\documentclass[12pt]{article}

\usepackage{graphics,cite,amssymb,epsfig,float,psfrag}
\usepackage[usenames,dvips]{color}
\usepackage{rotating}

\newcommand{\lsim}{\raisebox{-0.13cm}{~\shortstack{$<$ \\[-0.07cm] $\sim$}}~} 
\newcommand{\gsim}{\raisebox{-0.13cm}{~\shortstack{$>$ \\[-0.07cm] $\sim$}}~} 
 
\newcommand{\tb}{\tan\beta} 
\newcommand{\cotan}{{\rm cot}}
\newcommand{\beq}{\begin{eqnarray}} 
\newcommand{\eeq}{\end{eqnarray}}

\begin{document}

\vspace{1cm}



\vspace*{.5cm}

\begin{center}

{\large\bf The Left-Right asymmetry of top quarks in associated top--charged}

\vspace{.1cm}

{\large\bf Higgs bosons at the LHC as a probe of the tan$\beta$ parameter}

\vspace*{.8cm}







\mbox{\large   J. Baglio$^1$, M. Beccaria$^2$,  A. Djouadi$^{1,3}$  G. Macorini$^{2,4}$,}

\vspace*{2mm}

{\large   E. Mirabella$^5$, N. Orlando$^2$,} 

\vspace*{2mm}

{\large F.M. Renard$^3$, C. Verzegnassi$^4$,} 

\vspace*{.8cm}
$^1$ Laboratoire de Physique Th\'eorique, Universit\'e Paris XI et CNRS,
 Orsay, France.\\
$^2$ Dipartimento di Fisica, Universita del Salento and  INFN, Sezione di Lecce,
 Italy. \\
$^3$ Theory Unit, CERN, 1211  Gen\`eve 23, Switzerland.\\
$^4$ Niels Bohr International Academy and Discovery Center,
Blegdamsvej 17 DK-2100 Copenhagen, Denmark \\
$^5$ Institut de Physique Th\'eorique, CEA--Saclay, F--91191 Gif sur Yvette,
 France.\\
$^6$ Laboratoire Univers et Particules, U. Montpellier II,
France. \\
$^7$ Dipartimento di Fisica Teorica, Universita di Trieste and INFN Sezione di 
Trieste, Italy. \\

\end{center}

\vspace{1.4cm}

\begin{abstract} 
Many extensions of the Standard Model involve two Higgs doublet fields to break
the electroweak symmetry, leading to the existence of three neutral and two 
charged Higgs particles. In particular, this is the case of the Minimal
Supersymmetric extension of the Standard Model, the MSSM. A very important 
parameter is  $\tan\beta$ defined as  the ratio of the  vacuum 
expectation value of the two Higgs doublets. In this paper we focus on
the  left-right asymmetry  in the production of polarised top
quarks in association with charged Higgs bosons at the LHC. This quantity  
allows for a theoretically clean determination of  $\tan\beta$.  In the MSSM, 
the asymmetry remains sensitive to the  strong and electroweak 
radiative corrections and, thus, to the superparticle spectrum. Some possible 
implications of these results are discussed. 
\end{abstract} 

\thispagestyle{empty}
\setcounter{page}{0}
\newpage

\subsection*{1. Introduction} 

A widely studied extensions of the Standard Model (SM)  are the
 two--Higgs doublet model (2HDM) in which two SU(2) doublets of
complex scalar fields are introduced to break the electroweak symmetry
\cite{HHG}.  In particular the Higgs sector of the  
Minimal Supersymmetric extension of the Standard Model (MSSM)~\cite{Review} 
is a type II  2HDM.
These models lead to the existence of five scalar particles, two CP--even
bosons ($h,H$) a CP--odd one ($A$)  and two charged
particles ($H^\pm$). The Higgs sector of a 2HDM model is described by
six parameters. They can be chosen to be the four masses of the Higgs particles,   
the  mixing angle $\alpha$ in the CP--even Higgs sector and the ratio $\tb$ of the
vacuum expectation values of the two Higgs doublets. In the MSSM these parameters
are no longer independent.  The two parameters describing the Higgs sector 
of the MSSM  may be taken to be the charged Higgs mass $M_{H^\pm}$ and $\tb$.
The precise determination of   these parameters is  of great importance to 
identify the  underlying model and to determine its basic features.

Once the  Higgs bosons have been 
produced, their  mass can be measured  looking at the
kinematical distributions of the decays products~\cite{TDRs}. 
In the MSSM the parameter  $\tan\beta$ can be 
determined  looking at the total cross
section of processes involving Higgs bosons. For instance
in the MSSM 
the total cross sections $p p(\bar p) \to H,\, A$ are 
proportional to $\tan^2\beta$~\cite{gg+bbH}.
%
A measurement of the relevant production cross sections at the LHC, allows for a 
determination of $\tb$~\cite{Sasha-tb} with an uncertainty of the order of $30
\%$.%

Another interesting process  is the production of the charged
Higgs boson in association with  a top quark in bottom--gluon fusion 
at hadron  colliders~\cite{H+tbBorn,H+tbQCD,H+review,ClaudioH+,SaurabH+}
\beq 
bg \to t H^-, \qquad  \bar bg \to \bar t H^+,
\label{Eq:processes}
\eeq
in which the bottom quark is directly taken from the proton in a five 
flavor scheme. The cross section of this process is proportional  to the
square of the  Yukawa coupling $g_{H^\pm tb}$.  In 
type II 2HDMs  $g_{H^\pm tb}$ reads as follows~\cite{HHG}, 
\beq
g_{\rm H^\pm tb}&=& \frac{g}{\sqrt{2}M_W} V_{tb} \, \left\{ H^+ \bar{t} 
\left [m_b \tb\, P_R + m_t {\rm cot}\beta\, P_L \right ] b + {\rm h.c.} 
\right\},   
\label{coupling}
\eeq
where $g=e/s_W$ is the SU(2) coupling and $P_{L/R}= (1\mp \gamma_5) / 2$  are the
chiral projectors.  The Cabbibo--Kobayashi--Maskawa matrix element $V_{tb}$ can be set, to a good
approximation, to unity \cite{PDG}. At tree-level the total production cross sections  of the
processes in eq.~(\ref{Eq:processes})  
are equal and  proportional to $(m_t^2 \cot^2\beta+ m_b^2\tan^2\beta)$.  They are  
significant both in the  $\tb \le 1$ and in the $\tb \gg 1$ regions\footnote{The total cross section exhibits  a 
minimum at $\tan \beta = \sqrt{m_t /m_b} \approx 7$.}. 
In type I 2HDMs, all fermions couple to only one Higgs field. The   $g_{\rm H^\pm tb}$ coupling  has to be modified 
performing  the substitution $m_b \tb \to m_b \cot \beta$  in eq.~(\ref{coupling}).
The sum of the total corss section of the two processes in eq.~(\ref{Eq:processes}) 
is proportional to ${\rm cot}^2\beta$ and is enhanced for  small $\tb$ values only\footnote{In the MSSM 
the lower bound of the mass of $h$   requires that $\tb \gsim
2$--$3$~\cite{Review,PDG}. 
In a general  2HDM  $\tb$ is less constrained. The region $0.2 \lsim  \tb \lsim 50$ 
 is not ruled out and preserves  the perturbativity of the Higgs Yukawa
coupling~(2).}.

Besides the experimental uncertainties, 
the cross section measurement is plagued with various
theoretical uncertainties~\cite{Hpaper}.  The most important
uncertainties are related to the dependence of the observables 
on the renormalisation and factorisation scales, as  
well as the dependence on the choice 
of the parton distribution functions
(PDFs), and the related errors on the strong coupling constant $\alpha_s$. These
theoretical uncertainties can be of the  order of $20-30\%$~\cite{Sasha-tb}
and are a  major source of error in the determination of
$\tb$ directly from the Higgs production  cross section.

In this letter, we propose an alternative way to measure the parameter $\tb$
which is free of these theoretical uncertainties. The method uses the  left--right asymmetry  
constructed from the longitudinal polarisation of the top quarks 
produced in association with the charged Higgs bosons, the latter decaying via 
the clean and detectable  $H^\pm  \to \tau^\pm \nu$  decay channel. 
The polarisation asymmetry $A_{LR}^t $ is defined as the difference of cross sections
for the production of left--handed  and right--handed top quarks divided by their
sum\footnote{This asymmetry shares common interesting features with the long
celebrated $A_{LR}$ asymmetry for fermion pair production in longitudinally
polarized electron--positron annihilation on the $Z$ pole \cite{ALR}.}  
\beq 
A_{LR}^{t}
& \equiv & \frac{\sigma_L -\sigma_R} 
{\sigma_L+\sigma_R},
\label{Eq:ALRt}
\eeq  
where $\sigma_{L/R}$ is the total hadronic cross section of the process
of $t_{L/R}H^-$ associated production. The asymmetry is a ratio of observables 
of similar nature. Compared to the cross section, 
the asymmetry  is thus significantly  less affected by the scale and PDF uncertainties.
One is then mainly left only with the experimental
uncertainties in the determination  of the cross sections and with the measurement of
the polarisation of the top  quarks\footnote{We will not address here the issue
of the experimental determination of the top quark polarisation from analyses of
kinematical distributions of its decay products. For a detailed discussion, see
for instance, Ref.~\cite{Saurab}.}. In the MSSM, the asymmetry will nevertheless
remain sensitive to  the electroweak and strong radiative corrections from
supersymmetric particles which also strongly affect the cross sections at high
$\tb$ values \cite{H+tbQCD,DRweak,Deltab}. 

The polarisation asymmetry in $tH^- $ associated   production has been discussed in
Ref.~\cite{Claudio}, following an original study of the asymmetry in the case
of associated top--charged slepton  production in the MSSM \cite{Katri}. A
detailed  analysis of the top polarisation in $bg \to tH^-$ production has also
been given in Ref.~\cite{SaurabH+} which provides material that partly overlaps
with the one presented here.

In the next section, we discuss this asymmetry in the Born approximation and 
exhibit its dependence on $\tb$.  In section 3, we show that  it is essentially
independent of the scale and PDF choices but remains dependent on the important
SUSY radiative corrections that occur in the MSSM. A brief conclusion is given
in section 4.

\subsection*{2. The $\mathbf{A_{LR}^t}$ asymmetry at tree--level}

The starting point is the partonic process
\beq
b(p_b, \lambda_b) \; g(p_g, \lambda_g) \; \to \;  t(p_t,\lambda_t) \; H^-(p_H).
\label{Eq:process2}
\eeq
The momentum (helicity) of the particle $i$ is denoted by $p_i$ ($\lambda_i$).
In the Born approximation the process is mediated  by two Feynman
diagrams, one with $s$--channel bottom quark exchange and another with
$u$--channel top  quark exchange. In the case of type II 2HDM couplings   the helicity amplitude  
$F_{\lambda_b\lambda_g\lambda_t}$ reads as follows~\cite{ClaudioH+}
\beq
F_{\lambda_b\lambda_g\lambda_t} &=&
{g g_s \lambda^l \sqrt{x_+} \over2  M_W} \Bigg \{  
\frac{\delta_{\lambda_b\lambda_g}}{\sqrt{\hat s}} 
\left [ \lambda(1-r_t)s_{\theta / 2} \delta_{\lambda_b\lambda_t} 
+{1+r_t\over2} c_{\theta/2}
\delta_{\lambda_b-\lambda_t} \right ]  \nonumber \\
&+&  
\frac{m_t\delta_{\lambda_b\lambda_g}}{\hat u-m^2_t} \left [ (1+r_t) s_{\theta/2}
\lambda\delta_{\lambda_b\lambda_t} + {1-r_t\over2}
c_{\theta/2}\delta_{\lambda_b-\lambda_t} \right] \nonumber\\
&+& \frac{(1-r_t) s_{\theta / 2} \lambda 
\delta_{\lambda_b\lambda_t} }{ \hat u-m^2_t}\left [-p(1+c_\theta)
\delta_{\lambda_b-\lambda_g}+ d_t
\delta_{\lambda_b\lambda_g} \right ]\nonumber\\
&+&\frac{(1+r_t)  c_{\theta/2}
\delta_{\lambda_b-\lambda_t}}{2 (\hat u-m^2_t)}\left [p (1-c_\theta)
\delta_{\lambda_b-\lambda_g}+ d_t
\delta_{\lambda_b\lambda_g}\right ]\Bigg\}
\left [m_t \cot\beta \delta_{\lambda_tL}+m_b \tan\beta \delta_{\lambda_tR} \right]. 
\eeq
The partonic Mandelstam variables are defined as
$\hat s = (p_b+p_g)^2$ and   $\hat u=(p_b-p_H)^2$. 
 The angle $\theta$ is the azimuthal  angle in the center-of-mass frame,
while  $g_s$ is the strong coupling constant. The abbreviations
$d_t$, $r_t$,  $x_{\pm}$ and $\lambda$ read as follows
\beq
d_t = \sqrt{\hat s} -E_t+p \cos\theta, \qquad
r_t =  \sqrt{  \frac{x_-}{x_+} }. \qquad
x_{\pm} = \left (\sqrt{\hat s} \pm m_t \right ) ^2-M^2_{H^\pm},
\eeq
\beq
\lambda\!=\! \sqrt{
\left (1\!-\!(x_t\!+\!x_h  )^2 \right ) \left (1\!-\!(x_t\! - \!x_h)^2 \right )}, \nonumber
\eeq
while $p \equiv |{\bf p}_t|$, $c_\alpha \equiv \cos \alpha$, and $s_\alpha \equiv \sin \alpha$.
The partonic  cross sections for  L/R  polarized top quarks in the final state is 
\beq 
\hat \sigma_{L/R} =  \frac{p} {384 \pi \hat s^{3/2}}
  \int_{-1}^{+1} {\rm d}\!\cos \theta  \sum_{\lambda_b,\lambda_g} \big | F_{\lambda_b\lambda_g L/R} \big |^2.
\eeq
The integration over the angle $\theta$ leads to
\beq
\hat \sigma_{L} & = & \displaystyle \frac{G_F\alpha_s}{24\sqrt{2}
  \hat{s} \lambda}  \Bigg  \{   \lambda  \left [  m_t^2 \cotan^2\beta
   \left(\frac{7}{2} \lambda x_{ht}^2+2 x_{ht}^2+2
   \left(1-x_{ht}^2\right)^2+\frac{3}{2} (\lambda -1) \lambda
   \right) \right.   \nonumber \\
& & \displaystyle \left. - m_b^2 \tan^2\beta
    \left(-\frac{7}{2} \lambda x_{ht}^2+2 x_{ht}^2+2
   \left(1-x_{ht}^2\right)^2+\frac{3}{2} \lambda  (\lambda
   +1)\right) \right] + \nonumber \\ 
& & \displaystyle \Lambda  \Bigg[m_t^2 \cotan^2\beta \left(\left(x_{ht}^2+2 \lambda
    \right) \left(1-x_{ht}^2\right)^2+(\lambda +1) \left(x_{ht}^2 (\lambda
   +1)-1\right)\right)  \nonumber\\
& & \displaystyle + m_b^2 \tan^2\beta \left(\left(2 \lambda -x_{ht}^2\right)
   \left(1-x_{ht}^2\right)^2+\left((\lambda -1) x_{ht}^2+1\right)
   (1-\lambda )\right) \Bigg] \Bigg \}, \nonumber\\
\hat \sigma_{R} & = & \displaystyle \frac{G_F\alpha_s}{24\sqrt{2}
  \hat{s} \lambda} \left \{ \lambda  \left[ m_b^2 \tan^2\beta
   \left(\frac{7}{2} \lambda x_{ht}^2+2 x_{ht}^2+2
   \left(1-x_{ht}^2\right)^2+\frac{3}{2} (\lambda -1) \lambda
   \right) \right. \right. \nonumber \\
& & \displaystyle \left. - m_t^2 \cotan^2\beta
    \left(-\frac{7}{2} \lambda x_{ht}^2+2 x_{ht}^2+2
   \left(1-x_{ht}^2\right)^2+\frac{3}{2} \lambda  (\lambda
   +1)\right) \right] + \nonumber \\ 
& & \displaystyle \Lambda  \Bigg  [ m_b^2 \tan^2\beta \left(\left(x_{ht}^2+2 \lambda
    \right) \left(1-x_{ht}^2\right)^2+(\lambda +1) \left(x_{ht}^2 (\lambda
   +1)-1\right)\right)  \nonumber\\
& & \displaystyle  + m_t^2 \cotan^2\beta \left(\left(2 \lambda -x_{ht}^2\right)
   \left(1-x_{ht}^2\right)^2+\left((\lambda -1) x_{ht}^2+1\right)
   (1-\lambda )\right) \Bigg ] \Bigg \}.
\eeq
where $x_i = m_i/\sqrt{\hat s}$ and $x^2_{ht}=x^2_h  -x^2_t$ and $\Lambda$ is defined as 
\beq
\Lambda = \log \left (  \frac{1-x_{ht}^2+\lambda}{1-x_{ht}^2-\lambda
    } \right ).
\eeq
The total partonic cross section is then simply the sum of the cross sections 
$\hat \sigma_L$ and $\hat \sigma_R$
\beq
\hat \sigma_{\rm tot} = \frac{G_F \alpha_s}{24 \sqrt{2} \hat s}
 \left (m_t^2 \cot^2\beta\! +\! m_b^2 \tan^2 \beta \right ) 
\bigg\{  2 \left[1\!-\!2 x^2_{ht} 
(1\!-\!x^2_{ht})\right]\Lambda \!-\!  (3\!-\!7 x^2_{ht})  \lambda \bigg\}.  \ \ 
\label{siggbH+}
\eeq
As usual, these partonic cross sections have to be folded  with the
bottom--quark and gluon  densities to obtain the hadronic ones $\sigma_L,
\sigma_R$,  and $\sigma_{\rm tot}$. The expressions  in the type I 2HDM can be obtained
performing the substitution $m_b \tb \to m_b {\rm cot} \beta$.

\begin{figure}[!h]
\begin{center}
\mbox{
\epsfig{file=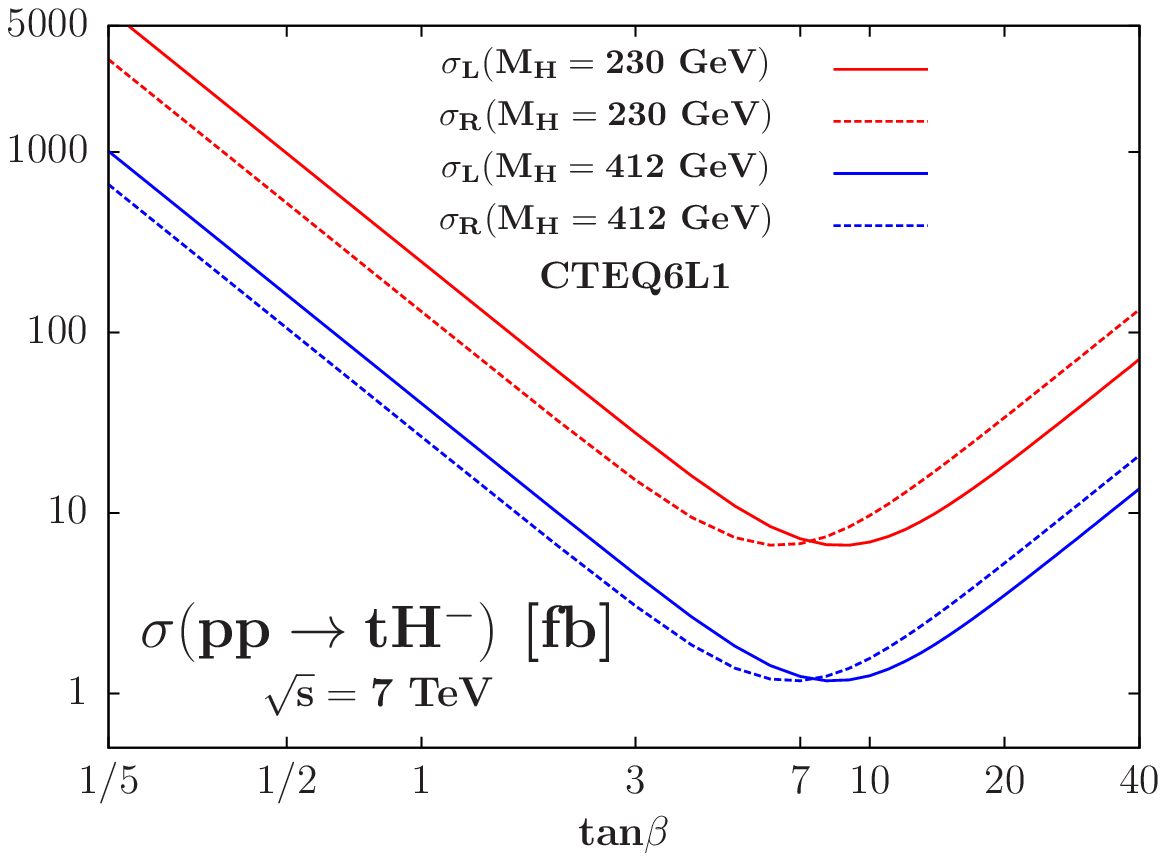,width=7.9cm}\hspace*{5mm}
\epsfig{file=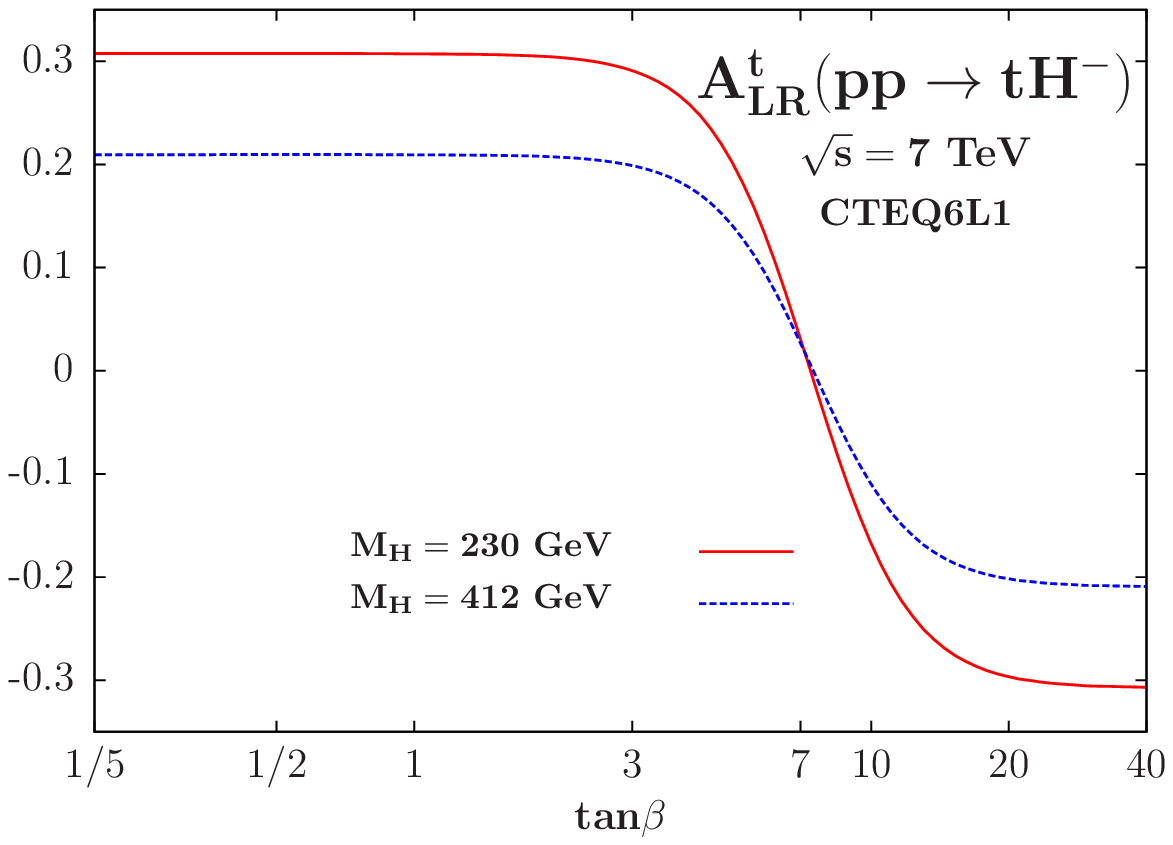,width=7.9cm} }
\end{center}
\vspace*{-6mm}
\caption[]{The cross sections $\sigma_L$ and $\sigma_R$ (left)
and the asymmetry $A_{LR}$ (right) at leading order in type  II 2HDMs as a  
function of $\tb$ in two benchmark scenarios with $M_{H^\pm}=230$ and $412$ GeV.} 
\vspace*{-2mm}
\label{typeII}
\end{figure}

\medskip
In Fig.~\ref{typeII} we display the left-- and right-- handed cross sections
$\sigma_L$ and  $\sigma_R$ as well as the asymmetry $A_{LR}^t$ at
the LHC as a function of $\tb$. We choose   two values of
 $M_{H^\pm}$, $M_{H^\pm}=230$ and $412$ GeV  corresponding to the  
two  2HDMs scenarios of type II  proposed  in
Refs.~\cite{DRweak} (LS2)  and \cite{Snowmass} (SPS1a) respectively.  The hadronic
center-of-mass  energy is
fixed to $\sqrt s=7$ TeV,  and we adopt the CTEQ6L1 leading order PDFs \cite{CTEQ}
with $\alpha_s(M_Z^2)=0.130$.
The  factorisation scale $\mu_F$ has  been set to the value
$\mu_0= (M_{H^\pm} +m_t)/6$ which
minimizes the higher order  QCD corrections~\cite{H+tbQCD}.
For the $H^-tb$
coupling, we use the on--shell top mass value $m_t=173.1$ GeV and  the
%
%
$\overline{\rm MS}$ mass of the bottom quark evaluated  at a scale of  $\mu =
\mu_F$; in the SUSY scenario analyzed in this paper $ m_b(\mu_F)$
ranges from 2.95 GeV to 3.10 GeV for all the values of $\mu_F$ considered.
%
%

As can bee seen $\sigma_L$ and $\sigma_R$ have the same order of magnitude:
they are large at small $\tb$ values, when the component $m_t\cot\beta$  of the
$H^-tb$ coupling is significant, as well as at large $\tb$ value when the
$m_b\tb$ component of the coupling is enhanced. The cross sections are equal and
minimal at the value $\tan \beta =\sqrt{m_t /m_b} \simeq 7$ for which the 
$H^-tb$ coupling is the smallest. Therefore  in type II 2HDM $A_{LR}^t$ is maximal at low 
$\tb$ values when the associated top quark is mostly left--handed  and minimal at large $\tb$
values when the top quarks are right handed.  For a given value of the charged
Higgs mass,  the modulus of $A_{LR}^{t}$ is the same in the  $\tb \gg 1$ and  in the
$\tb \le 1$ region. In the scenarios under consideration  $|A_{LR}^t|=0.31\; (0.21)$ for 
$M_{H^\pm}= 230 \;(412)$~GeV. The two $\tb$ regions differ
for the sign of the asymmetry.  Therefore the  sign of  $A_{LR}^{t}$   differentiates 
between the low and large $\tb$ scenarios. In the intermediate $\tb$ region, 
$\tan \beta  \simeq 7$ for which $\sigma_L \simeq \sigma_R$,
the asymmetry goes through zero. 

%
%
%
In a type I 2HDM,
the left-- and right-- components of the Yukawa coupling  $g_{H^\pm tb}$
 are both proportional to $\cot\beta$, and there is no $\tb$
dependence in $A_{LR}^t$. The asymmetry is thus constant
 and is simply given by the $A_{LR}^t$ value in
the corresponding type II model evaluated at  $\tb=1$.  For type I  2HDM characterized by 
 $M_{H^\pm}=230 \; (412)$~GeV the value of $A_{LR}^{t}$ can be 
 read off Fig.~\ref{typeII},  $A_{LR}^{t}=0.31 \; (0.21)$. Combining this value 
 with the value of $\sigma_{\rm tot} \propto \cot^2\beta$,  the predictions of 
 2HDMs of type I  and II can  eventually be discriminated. 

Note that while $\sigma_L, \sigma_R$ and thus $\sigma_{\rm tot}$ strongly depend
on the hadronic center-of-mass  energy, the asymmetry dependence of $A_{LR}^{t}$  is mild.
The asymmetry  is
comparable for $\sqrt s=7$ and $14$ TeV.  For instance at $\sqrt s = 14$ TeV
in the type I model one obtains $A_{LR}^t=0.27\;(0.18)$ for $M_{H^\pm}=230\;(412)$ GeV.

\subsection*{3. Scale and PDF dependence  and impact of  the NLO corrections}
\begin{figure}[!h]
\begin{center}
\mbox{
\epsfig{file=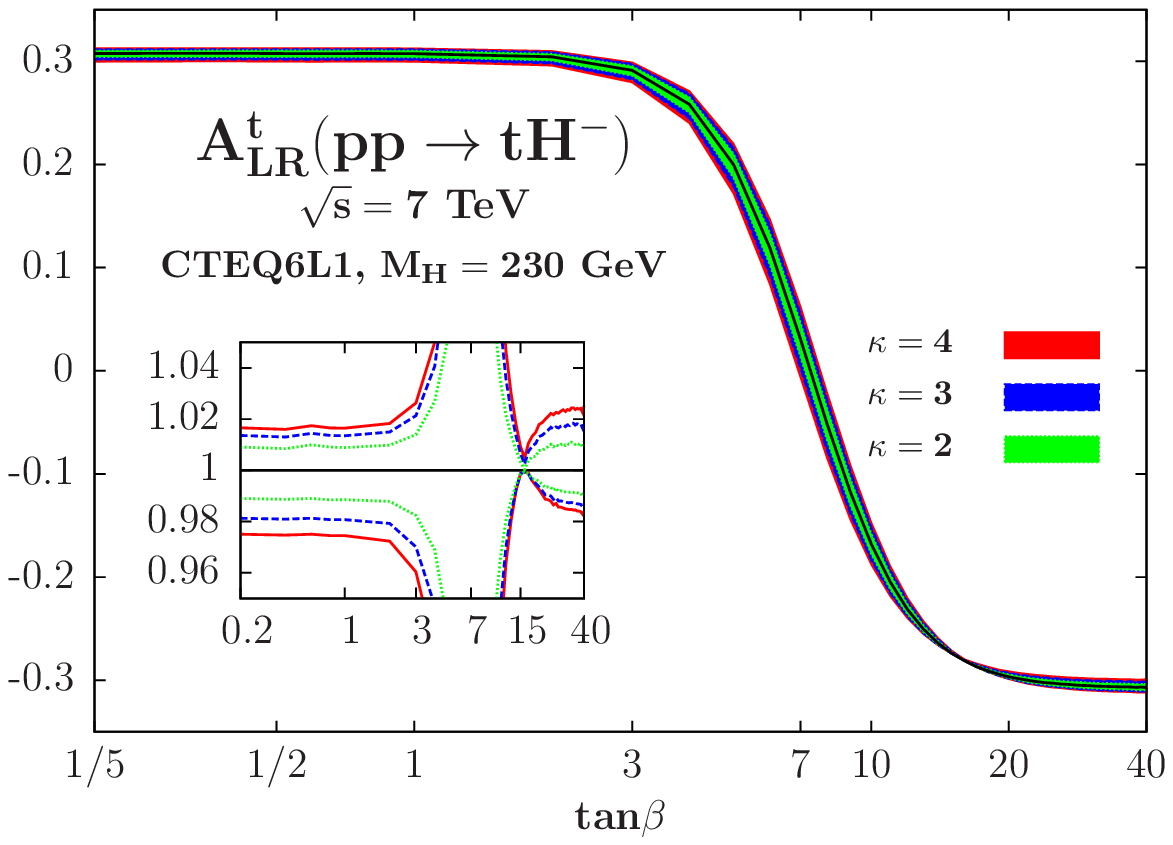,width=7.9cm}\hspace*{5mm}
\epsfig{file=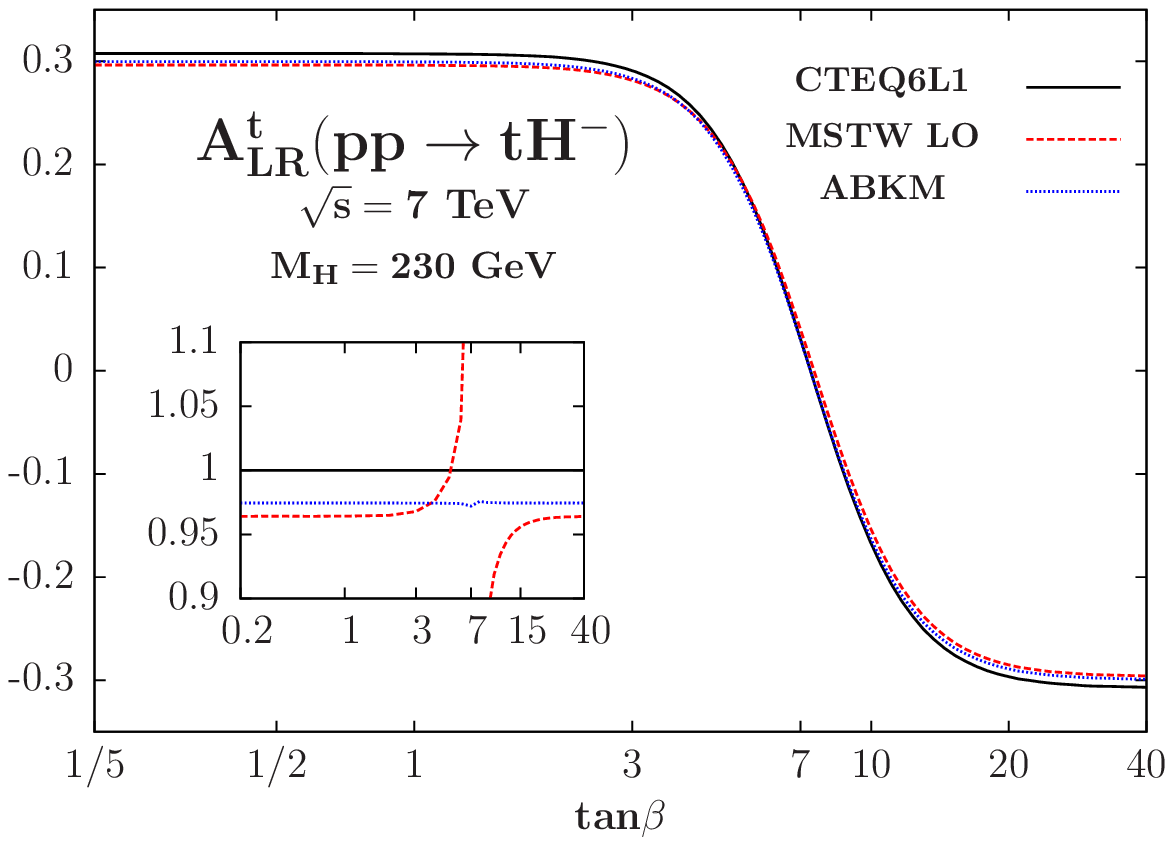,width=7.9cm} }
\end{center}
\vspace*{-6mm}
\caption[]{The scale variation (left) and the PDF dependence (right) of  the 
asymmetry $A_{LR}^t$ at leading order at the LHC with $\sqrt s=7$ TeV  
as a function of $\tb$. We consider the type  II 
2HDM  characterized by  $M_{H^\pm}=230$ GeV. In the
inserts, shown are the variations with respect to the central value.}
\vspace*{-2mm}
\label{scale}
\end{figure}

In this paper, the asymmetry $A_{LR}^t$ has been evaluated  at  tree--level. 
The yet uncalculated higher order QCD 
contributions on this observable can be estimated from its  dependence
on the  factorisation scale $\mu_F$ at which the process is evaluated.
Starting from our  reference  scale $\mu_0$ we  vary $\mu_F$ 
within the range $\mu_0/\kappa \!\le \!  \mu_F \!\le \! \kappa \mu_0$ with
the constant factor chosen to be $\kappa\!=\!2,3$ or $4$.
The left panel  of Fig.~\ref{scale} shows the variation of the polarisation 
asymmetry for the choices $\kappa=2,3$ and $4$. The insert 
shows the scale variation relative to the asymmetry 
value when the central scale is adopted. As  one can see
the scale dependence  is very mild. In   the  low and in the high $\tb$ region, it is at most at
the level of $2\%$, even for  $\kappa=4$.  At moderate
values of $\tb$,  $\tb \simeq 7$,  the relative variation
is  much larger since the asymmetry vanishes. However the absolute impact 
of the scale variation is comparable to the one obtained for 
low and high $\tb$ values, and thus small in absolute terms. I t is worth to notice
that the NLO QCD total cross section $\sigma_{\rm tot}$ exhibits a bigger
residual scale uncertainty estimated to be of the order of 10--20\% at
the LHC with $\sqrt s=7$ TeV~\cite{LHCXS}.

\medskip
Another source of uncertainty stems from the presently not satisfactory 
determination of the gluon and bottom quark PDFs. We estimate this type
of uncertainty evaluating  the asymmetry with several  PDF 
parameterizations.  In the  right panel  of Fig.~\ref{scale} we show the 
dependence of the asymmetry on $\tb$ when the  CTEQ,  the MSTW~\cite{MSTW},  and the  ABKM~\cite{ABKM} PDF 
sets are used. We consider the type II 2HDM characterized by $M_{H^\pm}=230$ GeV. As usual the asymmetry
has been computed at  the LHC with $\sqrt s=7$ TeV.  In the insert
we show the relative deviation from the CTEQ central prediction.
As can be seen,  the difference between the various predictions is rather small, less than 
few percents  at low and high $\tb$ values.  The peaks in the insert for 
$\tb \simeq 7$   correspond to the vanishing of $A_{LR}^t$.  The effect
of the PDF variation on the total cross section $\sigma_{\rm tot}$ is expected to be much 
larger. For instance at NLO  the  PDF uncertainty is expected to be of the order
of  $10\%$~\cite{LHCXS}.

\medskip
A final remark has to be made on the radiative corrections in
supersymmetric scenarios. In the MSSM, the process~(\ref{Eq:process2}) is affected
by radiative corrections involving the supersymmetric particle spectrum. 
The NLO QCD and electroweak corrections  have been discussed in 
Ref.~\cite{H+tbQCD} and in Ref.~\cite{DRweak} respectively.
Some of these corrections are known to be large  for high values of $\tb$ and
some other parameters such as the higgsino mass parameter $\mu$.
It turns out that the bulk of these radiative corrections can be accounted for 
modifying the Yukawa coupling~(\ref{coupling}) as described in 
Ref.~\cite{Deltab}.  
This modification  is equivalent of using an effective bottom--quark  mass. The approximation is rather good 
for the SUSY--QCD corrections (in particular  when the SUSY spectrum is rather heavy),
and slightly worse in the case of the electroweak ones.

\begin{figure}[!h]
\begin{center}
\vspace*{-2mm}
\mbox{
\epsfig{file=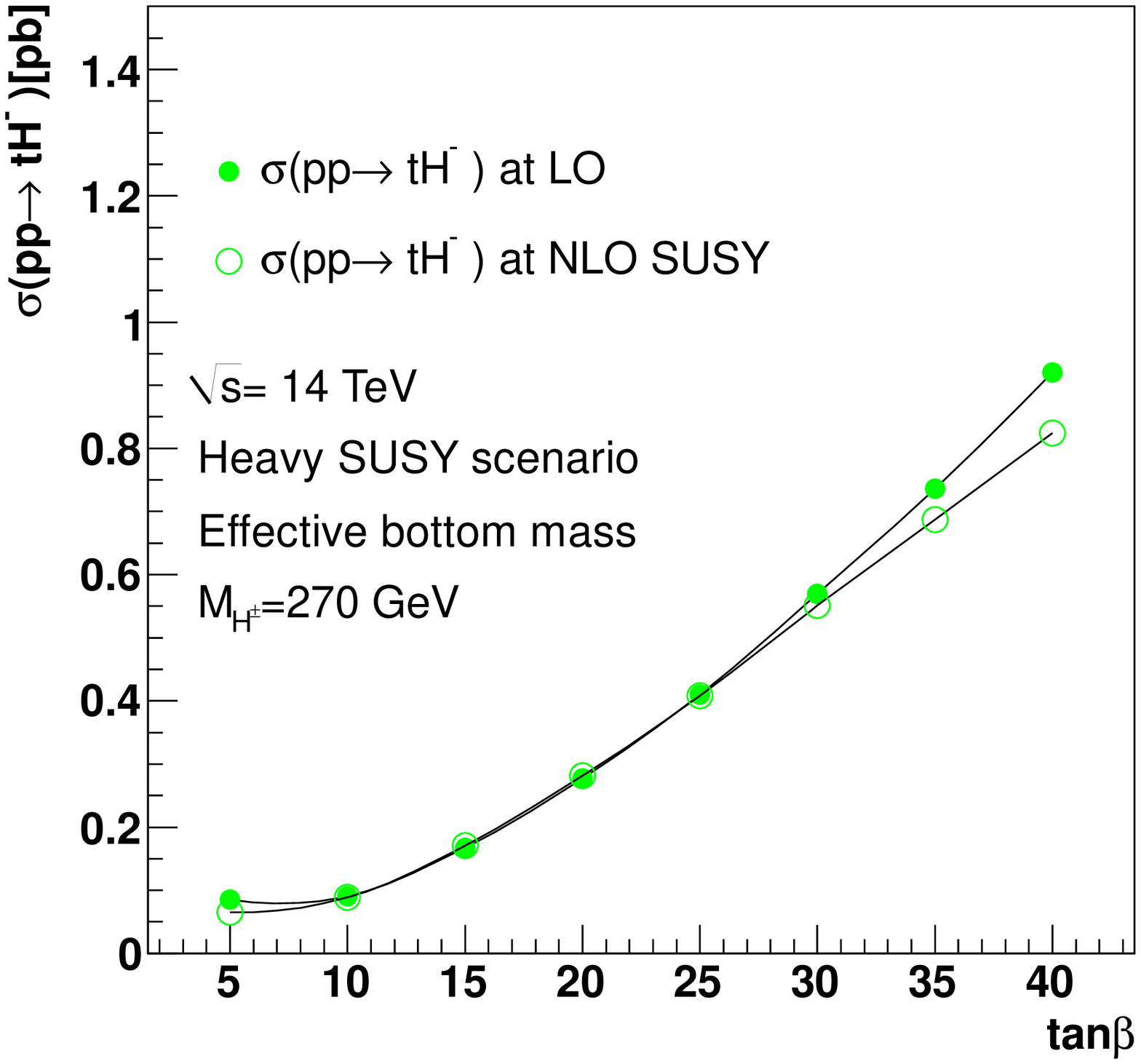,width=7.9cm} \hspace*{5mm}
\epsfig{file=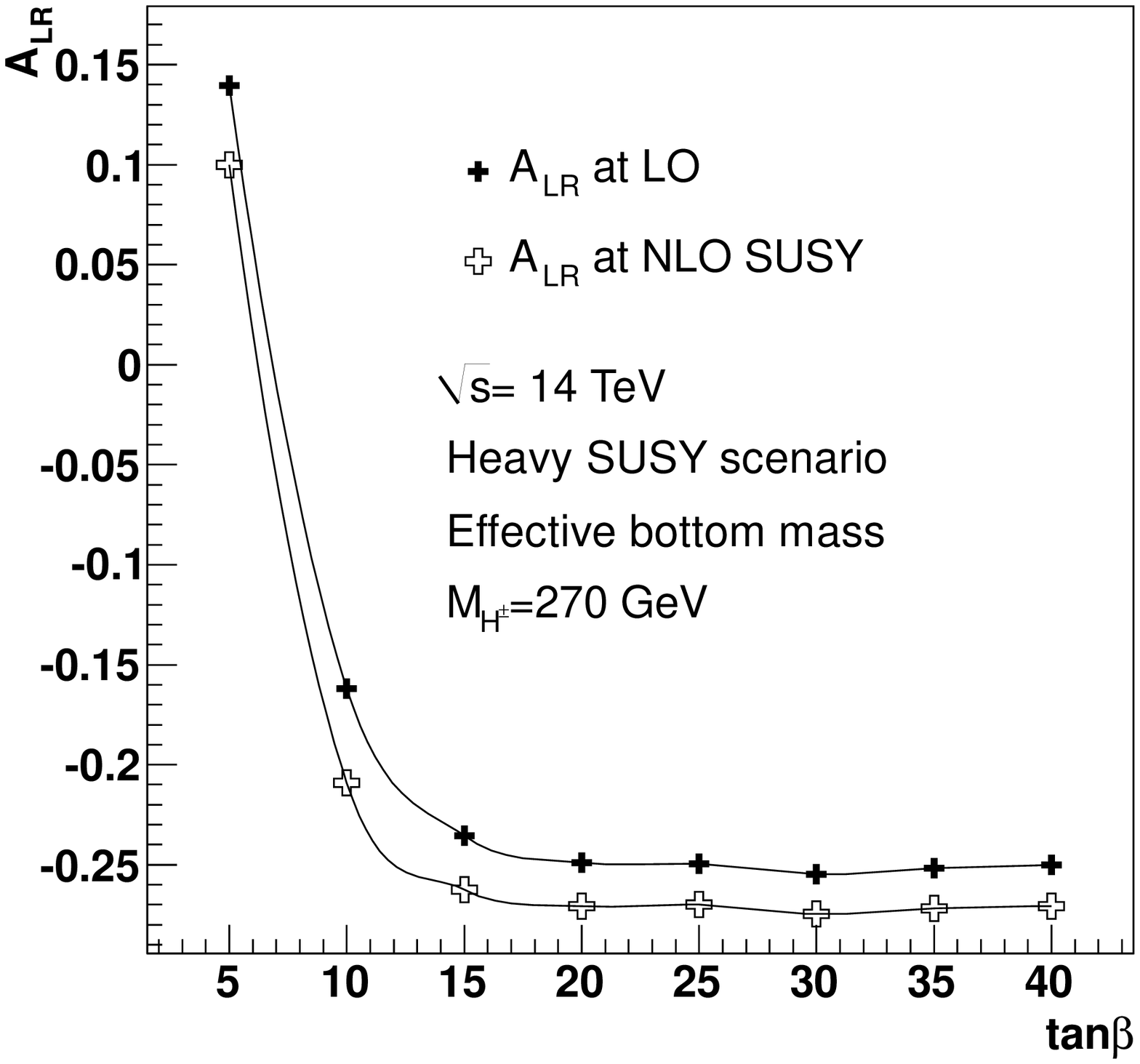,width=7.9cm}  }
\end{center}
\vspace*{-9mm}
\caption[]{The total production cross section  (left) and the asymmetry
$A_{LR}^t$ (right) at leading order and including the NLO  SUSY corrections
at the LHC with $\sqrt s=14$ TeV.  We consider the MSSM scenario 
of  Ref.~\cite{DRweak} characterized by  a heavy sparticle spectrum  
and $M_{H^\pm}=270$~GeV.  $\tb$ is varied from $5$ to $40$.}
\vspace*{-2mm}
\label{deltab}
\end{figure}

In Fig.~\ref{deltab}, we display the
impact of these NLO SUSY  radiative corrections within the MSSM   on both the total cross section
and the left--right asymmetry as a function of $\tb$. The other SUSY parameters are fixed according to the scenario
 presented in  Ref.~\cite{DRweak}, characterized by  a heavy superparticle spectrum and 
$M_{H^\pm}=270$ GeV. The 
SUSY QCD corrections are included in the  approximation of Ref.~\cite{Deltab}, 
while the electroweak and the (very small) QED corrections are computed exactly.
In the $\tb$ range considered the  approximation for the SUSY QCD contributions
 is expected to be valid.

As can be seen, the NLO  corrections  can be large  in both the cross section and the
asymmetry.  In the case of the latter observable the effect is of the order of $10\%$
in the  $\tb \ge 15$ region, where the asymmetry dependence  on $\tb$ is almost
flat. Therefore the 
asymmetry is  sensitive to the quantum contributions of the superparticle  spectrum.
A precise measurement of the asymmetry could allow to probe these additional
supersymmetric corrections and, hence, could help to discriminate between supersymmetric
and non--supersymmetric 2HDM of type II. 


\subsection*{4. Conclusion} 

In the process of $t H^-$ associated production at the LHC,  
the  left--right asymmetry, eq.~(\ref{Eq:ALRt}), obtained by 
identifying the polarisation of the top quarks  is rather  stable
against the scale and PDF variation. It  is still sensitive to quantum effects in new physics  scenarios
such as Supersymmetry. If measured  with some accuracy, the top quark
polarisation asymmetry in this process allows a very nice determination of the
parameter $\tb$. The combined measurement of the  production cross section and
the polarisation asymmetry could discriminate between various new physics
scenarios: two--Higgs doublet models of type I versus type  II and the MSSM
versus non-supersymmetric models, at least for intermediate values of $\tb$. 
For $\tb \gg 1$ or $\tb \le 1$ the method allows for the determination 
of the region of $\tan \beta$ but not for the  exact value of $\tb$, since in this two regions $A_{LR}^{t}$ has a plateau.
Note also that in the $\tb \le 1$ region the predictions for the asymmentry in the
THDM I and II coincide.
This polarisation asymmetry is thus worth
investigating theoretically and experimentally in more detail.

\bigskip

\paragraph {Acknowledgments.} 
Discussions with Rohini Godbole and Tilman Plehn are acknowledged.  The work of
A.D. is supported by the  ERC Grant ``Mass Hierarchy and Particle Physics at the
TeV Scale" and the project ANR CPV-LFV-LHC NT09-508531. E.M. is supported 
by the European Research Council under Advanced  Investigator Grant ERC-AdG-228301.

\end{document}